\documentclass[a4paper,12pt,english]{article}
\usepackage{amsfonts}
\usepackage{graphicx}
\usepackage{amsmath}
\usepackage{color}

\newcommand{\be}{\begin{equation}}
\newcommand{\ee}{\end{equation}}
\begin{document}

%\tableofcontents

\begin{titlepage}
\begin{center}

\noindent{{\LARGE{On string theory on AdS$_3\times {M}_7$ in the tensionless limit}}}

\smallskip
\smallskip

\smallskip
\smallskip
\smallskip
\smallskip
\smallskip
\smallskip
\noindent{\large{Gaston Giribet}}

\smallskip
\smallskip

\smallskip
\smallskip

\smallskip
\smallskip
\centerline{Physics Department, University of Buenos Aires FCEyN-UBA and IFIBA-CONICET}
\centerline{{\it Ciudad Universitaria, pabell\'on 1, 1428, Buenos Aires, Argentina.}}

\end{center}

\bigskip

\bigskip

\bigskip

\bigskip

\begin{abstract}
We review old and recent results on a special limit of string theory on AdS$_3\times M_7$ with pure NS-NS fluxes: the limit in which the string length $\ell_s=\sqrt{\alpha'}$ equals the AdS$_3$ radius $R $. At this point of the moduli space, the theory exhibits special properties, which we discuss. Special attention is focused on features of correlation functions that are related to the non-compactness of the boundary CFT target space, and on how these features change when the point $k\equiv R^2/\alpha ' =1$ is approached. Also, we briefly review recent proposals for exact realizations of AdS/CFT correspondence at this special point. 
\[\]    
This is the transcript of the talk delivered by the author at the 8$^{\text{th}}$ edition of the Quantum Gravity in the Southern Cone conference, held in Patagonia, December 16$^{\text{th}}$ - 20$^{\text{th}}$, 2019.

\end{abstract}
\end{titlepage}
%%%%%%%%%%%%%%%%%%%%%%%%%%%%%%%%%%%%%%%%%%%%%%%%%%%%%%%%%%%%%%%%%%%%%%%%

%\newpage

\section{Introduction}

Superstring theory on AdS$_3 \times M_7$ with pure NS-NS fluxes provides an excellent arena to investigate AdS/CFT correspondence in the stringy regime. On the one hand, the access to the worldsheet formulation makes it possible to compute 3-point scattering amplitudes explicitly and obtain expressions for finite $R^2/\alpha '$. On the other hand, the existence of non-renormalization theorems \cite{Verlinde} enables to compare such observables in the bulk with those in the boundary, showing a perfect agreement between protected correlation functions \cite{Pakman, Gaberdiel, Sever, Rastelli, Cardona}. Such non-renormalization theorems are crucial ingredients in the checks of AdS$_3$/CFT$_2$ correspondence, as the bulk and the boundary computations are usually performed at different points of the moduli space: While the worldsheet computation of 3-point functions of chiral states is done for the theory with pure NS-NS fluxes, the boundary correlators are computed at the so-called orbifold point. Still, the fusion rules and the structure constants of the $\mathcal{N}=2$ chiral ring in the bulk can be shown to exactly matches those in the symmetric product CFT$_2$ in the large $N$ limit.

Superstring theory admits supersymmetric backgrounds of the form AdS$_3 \times  S^3 \times  M_4$, with $M_4$ being $  T^4,$ $  K_3,$ or $S^3 \times S^1$. These backgrounds can be supported by either NS-NS or R-R fluxes. In the cases $M_4=  T^4$ and $M_4=  K_3$, the dual theory can be constructed in terms of a symmetric product on $M_4$, and this has been understood long time ago. In the case $M_4=S^3 \times S^1$, in contrast, the construction of the dual theory has shown to be more elusive.

Very recently, the interest in superstrings on AdS$_3 \times  S^3 \times  M_4$ has been renewed \cite{Gopakumar}, and very interesting new results have been obtained \cite{Gaberdiel:2017oqg, 1704.08667, 1803.04420, 1803.04423, 1812.01007, 1904.01585, 1903.00421, 1907.13144, otros, otros2}, especially in the context of AdS$_3$/CFT$_2$. Much of the attention in the recent studies is focused on the special limit in which the string length $\sqrt{\alpha'}$ equals $R$, the radius of AdS$_3$ space. This can be thought of as a sort of ``loose string limit'', in which the theory exhibits special properties such as symmetry enhancement, gapless continuous spectrum, new high energy phases, and others. Among the most salient properties that the theory seems to have at $k\equiv R^2/\alpha' =1$, we find that it admits a remarkably simple CFT$_2$ holographic dual. A series of recent papers investigates AdS$_3$/CFT$_2$ correspondence at the stringy point $k=1$, including not only the case $M_4 = T^4$ but also the case $M_4 = S^3\times S^1$. Here, our aim is to review the special properties that the theory exhibits at that point and summarize the recent investigations on its holographic description. 

In section 2, we will review string theory on AdS$_3$ from the worldsheet perspective. We will review the spectrum in terms of unitary representations of $SL(2,\mathbb{R})_k$, as constructed by Maldacena and Ooguri in references \cite{MO1, MO2}. In section 3, we will mention the special values of $k$ for which the theory exhibits special features. In sections 4, 5, and 6, we will focus our attention to the special value $k=3$ and study the properties that the bosonic theory presents at that point. In section 7, we turn our attention to the superstring theory at $k=1$, which is the supersymmetric analog of the bosonic $k=3$. We review recent attempts to construct a CFT$_2$ dual for the superstring theory on AdS$_3\times S^3 \times  M_4$, NS-NS backgrounds in the stringy regime.

\section{String theory on AdS$_3$}

Let us first consider the bosonic theory on AdS$_3\times M$. The string $\sigma $-model on AdS$_3$ with NS-NS fluxes is defined by the Polyakov action
\begin{equation}
S_{\text{string}}=\frac{1}{2\pi \alpha'}\int_{\Sigma } d^2z \ \big(\sqrt{h}h^{\alpha \beta} G_{\mu \nu}(X) + \epsilon^{\alpha \beta} B_{\mu \nu}(X)\,\big)\partial_{\alpha }X^{\mu} \partial_{\beta }X^{\nu}\, ,
\end{equation}
where we are using standard self-explanatory notation. It turns out that this action coincides with that of the WZW model for the non-compact group $SL(2,\mathbb{R})$; namely
\begin{equation}
S_{\text{WZW}}=\frac{k}{2\pi }\int_{\Sigma } d^2z \ \delta^{\alpha \beta}\, \text{Tr}\big( g^{-1}\partial_{\alpha }g\,g^{-1}\partial_{\beta }g\big) +\frac{ik}{6\pi }\,\Gamma_{\text{WZ}} \ \ , \ \ g\in SL(2,\mathbb{R}).
\end{equation}
where $\Gamma_{\text{WZ}}$ stands for the Wess-Zumino term. The WZW level is given by $k=R^2/\alpha' $, so that the semiclassical limit corresponds to $k\to \infty$. Here we will be concerned with the opposite limit, namely $k\sim \mathcal{O}(1)$. 

The acion of the $SL(2,\mathbb{R})_k$ WZW model can be represented as follows
\begin{equation}\label{Waki}
S_{\text{WZW}}=\frac{1}{2\pi }\int_{\Sigma } d^2z \,\Big(\partial \phi \bar{\partial }\phi -\beta \bar{\partial }\gamma - \bar{\beta }{\partial }\bar{\gamma } -2\pi \lambda \beta \bar{\beta }e^{-\sqrt{\frac{2}{k-2}}\phi } \Big),
\end{equation}
where the scalar field $\phi $ parameterizes the radial direction in AdS$_3$. Fields $\gamma$, $\bar{\gamma}$ are coordinates in the boundary of AdS$_3$, which is located at $\phi =\infty$. Action (\ref{Waki}) already includes quantum (finite-$k$) corrections, which take the form of a linear dilaton term that provides the field $\phi$ with a background charge $1/\sqrt{k-2}$ that does not appear explicitly in the conformal gauge we have chosen to use here. $\lambda $ is then related to the string coupling, as it can be changed by shifting the linear dilaton. Quantum corrections also explain the exponent in the last term of (\ref{Waki}), which follows from a rescaling of $\phi $ that is required to canonically normalize the quantum corrected kinetic term. Fields $\beta $, $\bar{\beta}$ are auxiliary fields and, as such, can be integrated out and replaced back in the action. However, their presence makes the theory more tractable, as it appears as a perturbation of the free theory $\lambda =0$, the latter consisting of a $\beta$-$\gamma$ ghost system coupled to $\phi $. Representation (\ref{Waki}) will be particularly useful later. 

The $SL(2,\mathbb{R})_k$ WZW model exhibits $\hat{sl}(2)_k \oplus \hat{sl}(2)_k$ affine Kac-Moody symmetry. Thus, the stress tensor of the theory follows from the Sugawara construction. This yields the Virasoro central charge 
\begin{equation}
c=\frac{3k}{k-2}+c_{M}=26,
\end{equation} 
where $c_{M}$ is the contribution to the central charge coming from the internal manifold $M$.   

The Hilbert space of the theory on AdS$_3\times M$ is given by the direct product of states $\vert h, m, \omega \rangle \times \vert h, \bar{m}, \omega \rangle $ representing string states in AdS$_3$, and states $ \vert \Delta_{{M}},\bar{\Delta}_{{M}}\rangle $ representing the string configurations in the internal manifold $M$. The indices $h,\, m, \, \omega $ label unitary representations of the universal covering of $SL(2,\mathbb{R})_k$, including the so-called spectrally flowed representations \cite{MO1}. In terms of these variables, the worldsheet conformal dimensions $( \Delta,\bar{\Delta} )$ take the form
\begin{equation}\label{D1}
\Delta = \frac{h(1-h)}{k-2}-m\omega -\frac{k}{4}\omega^2+ N + \Delta _{{M}} = 1,
\end{equation}
\begin{equation}\label{D2}
\bar{\Delta} = \frac{h(1-h)}{k-2}-\bar m\omega -\frac{k}{4}\omega^2+ \bar N + \bar{\Delta}_{{M}}  = 1,
\end{equation}
where the Virasoro constraints have already been imposed. 

The physical interpretation of the indices $h,\, m,\, \bar{m},\, \omega,\, N,\, \bar{N}$ is the following: $h$ is associated to the radial momentum of strings; $\omega \in \mathbb{Z}$ is the winding number; $E\equiv m+\bar{m}+k\omega \in \mathbb{R}$ is the energy in AdS$_3$; $J\equiv m-\bar{m}\in \mathbb{Z}$ is the angular momentum; $N$, $\bar{N} \in \mathbb{Z}_{\geq 0}$ are the string excitation numbers; $\Delta_M$, $\bar{\Delta}_M$ are the contributions to the worldsheet conformal dimension coming from $M$; see \cite{MO1} for details.  

The unitary representations of $SL(2,\mathbb{R})_k$ that describe string states in AdS$_3$ have been identified in \cite{MO1}. These belong to the discrete lowest- and highest-weight series, and to the principal continuous series. In the case of the discrete series, the value of $h$ has to be additionally restricted: No ghost theorem demands $h$ to belong to the segment $0 <  h  < k/2$, while normalizability demands, additionally, $1/2 \leq  h  < k/2$. Finally, spectral flow symmetry demands the stronger constraint; namely \cite{MO1}
\begin{equation}\label{A1}
2h-1\, \in \, [0,k-2] .
\end{equation}
In the case of the continuous series, $h$ takes the non-real value
\begin{equation}\label{A2}
 2h-1\, \in \, i\mathbb{R} .
\end{equation}
Short strings in AdS$_3$ are described by the $SL(2,\mathbb{R})$ discrete series obeying (\ref{A1}), while long strings are described by $SL(2,\mathbb{R})$ continuous series (\ref{A2}). The latter string states constitute a continuous part of the energy spectrum, and they play a important role in the discussion of the dual CFT$_2$.

\section{Special point(s)}

As said, we will be concerned with $k={R^2}/{\alpha '}\sim \mathcal{O}(1)$. More precisely, we will study a special value of $k$ at which the theory exhibits peculiar properties. There are, in fact, more than one specific order-one value of $k$ for which the theory behaves in a special way. Probably the most evident one is $k=2$, which corresponds to a singular limit of the worldsheet CFT: At $k=2$ the first term in (\ref{D1})-(\ref{D2}) diverges, what reflects the fact that the Sugawara stress tensor is not well-defined when the Kac-Moody level equals the Coxeter number of the group. This singular point is relevant in the study of the geometric Langlands program \cite{Frenkel} and for the connection between the affine Kac-Moody symmetry and the $W$-symmetry. For the worldsheet CFT to be well-defined at $k=2$, a rescaling of certain physical quantities such as the stress tensor is needed in order to avoid infinities. The fact that this point is somehow special can be anticipated by observing that, at $k=2$, the discrete spectrum (\ref{A1}) collapses to the point $h=1/2$ where it fuses with the continuous spectrum (\ref{A2}). Also, when $k\to 2$ the central charge of the $SL(2,\mathbb{R})_k$ WZW model diverges, and the model relates in many ways with Liouville field theory in its classical limit $c_L\to \infty$.

Another interesting limit is ${k\to 0}$, in which the central charge of the $SL(2,\mathbb{R})_k$ WZW model vanishes. While its geometrical interpretation is elusive, as well as its Lagrangian realization, the worldsheet theory can still be studied in that limit, for example resorting to the connection to Liouville field theory via the so-called $H_3^+$ WZW-Liouville correspondence \cite{RibaultTeschner, HikidaSchomerus}. In fact, in the limit $k\to 0^+$ the $H_3^+ =SL(2,\mathbb{C})/SU(2)$ WZW $n$-point correlation functions reduce to $n$-point correltion functions of the analytically continued $c_L=-2$ Liouville field theory in a direct way \cite{Nakayama}, as the simplest degenerate fields coincide with the screening fields at that point. $k=0$ is also a fixed point of the so-called Fateev-Zamolodchikov-Zamolodchikov (FZZ) duality, connecting the gauged $SL(2,\mathbb{R})_k /U(1)$ WZW model to the so-called sine-Liouville CFT, a two-fields theory with an interaction potential of the form
\begin{equation}
V(\varphi , X) = e^{\sqrt{\frac{k-2}{2}}\varphi }\ \cos (\sqrt{k}X). 
\end{equation} 
FZZ duality also reduces to a simple identity at the point $k=0$, where the sine-Gordon field $X$ and the Liouville field $\varphi $ decouple. %As $k=0$ both the $H_3^+$ WZW-Liouville correspondence and the FZZ duality reduce to simple identities. 

A third point that is special is {$k=3$}, which is actually the point we will be concerned with here. A $k=3$, the bosonic theory exhibits higher-spin symmetry, phase transitions, gapless continuous spectrum, and special structure of its correlation functions. 
%It can also be seen that, when extended to the superstring, the theory at such special point appears to admit a remarkably simple dual description in terms of a symmetric product CFT$_2$. 
In the supersymmetric theory, the level gets shifted as $k\to k+2$ with respect to the bosonic theory; recall that $-2$ is the Coxeter number of $SL(2,\mathbb{R})$. More precisely, the $k$-dependent prefactor in the quadratic Casimir receives a finite-$k$ correction, and since here we are interested in the finite $k$ regime, such correction is important: The value $k=1$ is to the superstring theory on AdS$_3$ what the value $k=3$ is to its bosonic counterpart. Then, for simplicity, and since the main features we want to discuss are shared by the supersymmetric and the bosonic theories, we will first focus on the bosonic string theory at $k=3$, and later we will go back to the supersymmetric string at $k=1$. %Notice that the superstring backgrounds AdS$_3\times S^3\times M_4$ appear as a near horizon limit of the NS$_1$/NS$_5$ system, and $k=1$ corresponds to the setup with a single unit of NS flux.

\section{Long strings}

A special feature that string theory exhibits at the special point is a phase transition. While for $k>3$ the $SL(2,\mathbb{C})$ invariant vacuum of the boundary CFT$_2$ is normalizable and the generic high energy states correspond to black holes, for $k<3$ the behavior is qualitatively different: As shown in \cite{KS}, the vacuum of the boundary CFT$_2$ is associated to a worldsheet state with $h=1$, and when $k<3$ the unitarity bound (\ref{A1}) excludes such state. Then, as argued in \cite{Giveon:2005mi}, this indicates that the $SL(2,\mathbb{C})$ invariant vacuum as well as the states corresponding to BTZ black holes \cite{BTZ} are absent from the spectrum if $k<3$. As a consequence, the high energy states in the theory with $k<3$ are not black holes but rather correspond to long strings that extend to the boundary of AdS$_3$, where they become weakly coupled. The entropy of black holes coincides with that of long strings at $k=3$ \cite{Giveon:2005mi}, while for $k<3$ the effective central charge controlling the asymptotic growth of states starts to differ from the actual central charge $c=6k$ \cite{GKS, KS, Ooguri} of the boundary CFT$_2$. 

All this is associated to another phenomenon that occurs at $k=3$: In \cite{SeibergWitten}, Seiberg and Witten studied long fundamental strings in AdS$_3$ and argued that, at least at large $k$, a single long string is effectively described by a Liouville field theory with central charge
\begin{equation}
c_L=1+6Q^2\, , \nonumber
\end{equation}
and background charge
\begin{equation}
Q={\sqrt{k-2}}-\frac{1}{\sqrt{k-2}}\, ;\nonumber
\end{equation}
so that at large $k$ one gets $c_L\simeq 6k$. In this effective description, the Liouville field is associated to the radial distance of the long string from the boundary of AdS$_3$ and it parameterizes the transverse excitations. The normalizable states of Liouville theory are states with momentum $\alpha \in Q/2 +i \mathbb{R}$ and conformal dimension $ \Delta=\alpha (Q-\alpha )\in Q^2/4+\mathbb{R}_{\geq 0}$; thus, the theory has a threshold 
\begin{equation}
\Delta\geq \frac{Q^2}{4}= \frac 14  \frac{(k-3)^2}{k-2}\label{gap} 
\end{equation}
that vanishes when $k\to 3$. In other words, at the special point this sector becomes gapless. 

The vanishing of the gap (\ref{gap}) at $k=3$ has been recently discussed in relation to other phenomenon observed, namely the emergence of a plethora of light states at that point: It has been observed in \cite{Gaberdiel:2017oqg} that, at $k=3$, the bosonic theory contains in the spectrum massless higher-spin fields. Looking at the worldsheet formulae (\ref{D1})-(\ref{D2}), Hull, Gaberdiel and Gopakumar noticed in \cite{Gaberdiel:2017oqg} that there is an infinite stringy tower of massless higher-spin fields which are actually part of the continuum. These state appear at $k=3$, and correspond to the representation with $h=1/2$ in the spectral flow sector $\omega = 1$. The states are those with values $m+{k}\omega/{2} =m+3/{2} = N$, $\bar{m}+{k}\omega /{2}=\bar{m}+3/{2}=\bar{N}= 0$. This yields $E=m+\bar{m}+k\omega =m+\bar{m}+3 = N$ and $J=m-\bar{m} = N$; namely
\begin{equation}
E-J=0 \, , \ \ E+J = 2N
\end{equation}
with $N\in \mathbb{Z}_{\geq 0}$ being arbitrary. The fact that these states appear at $k=3$ rather than at the {\it naive} tensionless point $k=2$ was surprising at the beginning. It has been argued that this is associated to the disappearing of the massless
graviton $h=1, \, \omega =0$ when $k < 3$. Spectral flow identifies representations $h,\, \omega=0$ with representations $\hat{h}=k/2-h,\, \omega=\pm 1$, and at $k=3$ this relates the massless state $h=1,\, \omega =0$ to the states $\hat{h}=1/2, \, \omega=1$. %However, understanding the precise origin of these light states in physical terms still requires further investigation. 

In \cite{1704.08667} the worldsheet analysis of the higher-spin phenomenon was extended to the superstring theory. This was partially motivated by the analysis done for the bosonic case \cite{Gaberdiel:2017oqg} but also by the observation that the CFT$_2$ dual of superstrings on AdS$_3 \times S^3 \times T^4$, meaning the symmetric orbifold of $T^4$, contains a closed higher-spin subsector. In \cite{1704.08667}, the authors identified the states that make up the leading Regge trajectory, and they showed that such states accommodate into Vasiliev higher-spin theory. At $k=1$, such higher-spin states become massless and, consequently, the supersymmetric theory also contains a stringy tower of massless higher-spin fields which come from the long string sector.

\section{String interactions}

Perhaps the most interesting properties of the theory at $k=3$ have to do with those features of the string amplitudes in AdS$_3$ that are associated to the non-compactness of the boundary CFT$_2$ target space. One such property is related to the factorization of the $4$-point amplitude (see Fig. 1). As explained by Maldacena and Ooguri in \cite{MO3}, the OPE in the boundary CFT$_2$ is well-defined only when the following constraints are obeyed
\begin{equation}\label{La44}
h_1+h_2 < \frac{k+1}{2} \ ,\  \ h_3+h_4 < \frac{k+1}{2},
\end{equation}
where $h_i$ are the external momenta involved in the $4$-point WZW correlator. In general, (\ref{La44}) is more restrictive than the bounds on $h_1+h_2 $ and $h_3+h_4 $ imposed by unitarity. It was explained in \cite{MO3} that when (\ref{La44}) is not satisfied, then there appear in the intermediate channels of the $4$-point function contributions that do not have an interpretation as the exchange of
intermediate physical states. This signals a breakdown of the OPE, which does not mean that one has to include additional physical states
in the theory for the OPE to close; it rather means that when (\ref{La44}) is not obeyed then the OPE is not well-defined in the target space theory due to its non-compactness \cite{MO3}. However, something remarkable happens at $k=3$: As we discussed in section 2, unitarity demands (\ref{A1}) and thus
\begin{equation}
h_1+h_2 < {k-1} \ ,\  \ h_3+h_4 < {k-1}.\label{Segundona}
\end{equation}
Then, one might ask when does (\ref{Segundona}) strictly imply (\ref{La44}). The answer is that this only happens for $k\leq 3$. That is, for $k\leq 3$ it turns out that the unitarity bound does imply the factorization of the $4$-point function and no further restrictions are needed for the OPE to hold. 
\begin{figure}
\ \ \ \  \ \ \ \includegraphics[width=5.7in]{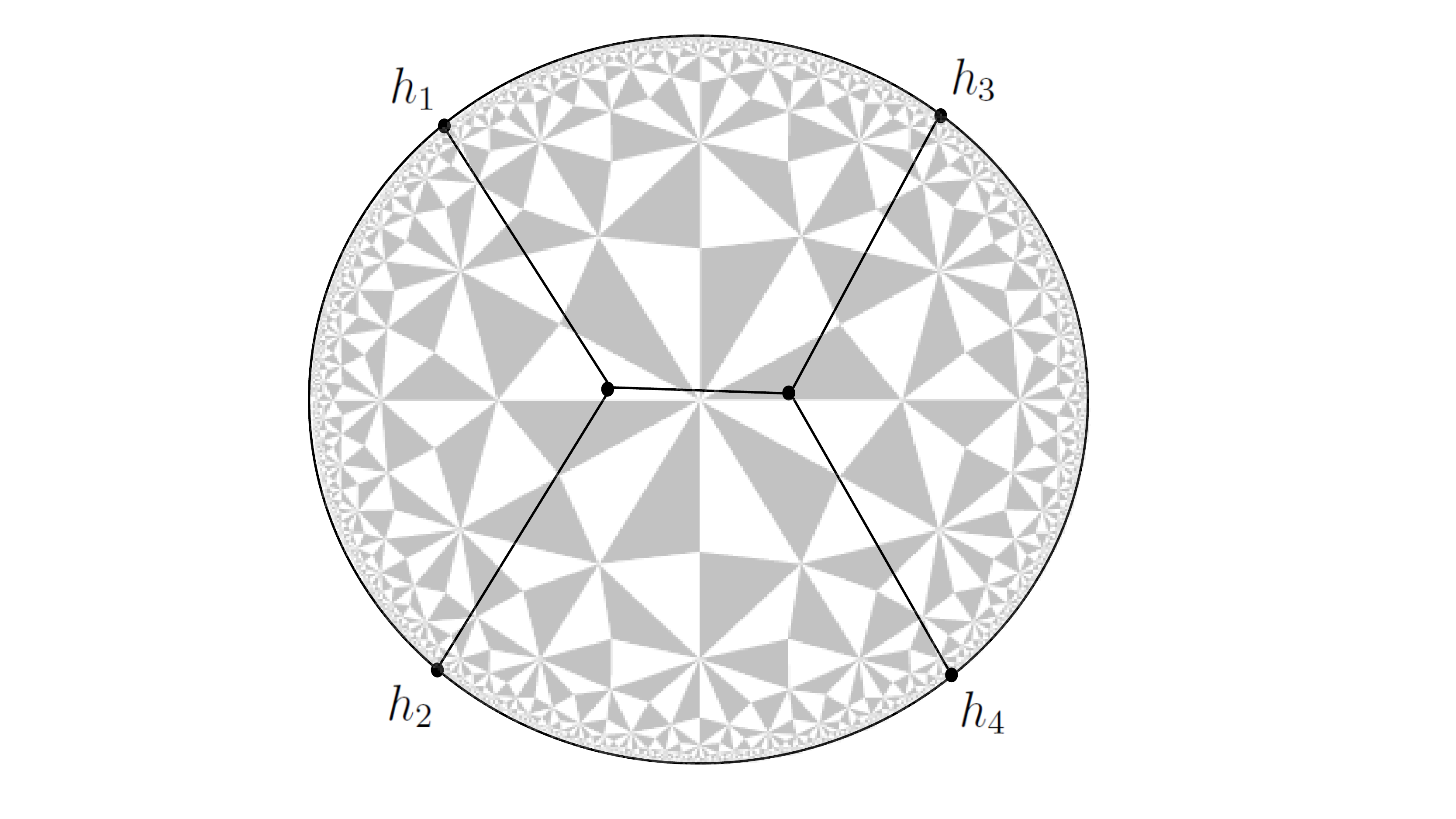}
\caption{Scheme of the $s$-channel factorization of the $4$-point function.}
\label{Figure1}
\end{figure}

\section{Worldsheet instantons}

A similar phenomenon occurs with certain non-perturbative worldsheet configurations: Also in \cite{MO3}, it was observed that the $n$-point function contains undesirable divergences that are due to the presence of worldsheet instantons. These correspond to non-local effects in the target space theory. The configurations responsible for this are holomorphic worldsheet solutions that can explore the non-compact space with no cost in the action (see Fig. 2). In the representation (\ref{Waki}), such solutions are, for example, those obeying $\bar{\partial } \gamma (z)=0$, for which the effective gravitational potential cancels out. To see this explicitly, one can integrate out in the auxiliary fields $\beta $, $\bar{\beta }$ in (\ref{Waki}) to get a potential term of the form
\begin{equation}
\int d^2z\, \partial \bar{\gamma} \bar{\partial} {\gamma}\, e^{\sqrt{\frac{2}{k-2}}\phi}
\end{equation}
which actually vanishes for holomorphic maps $\gamma(z)$. Such holomorphic solutions have been those considered in \cite{SeibergWitten} to describe classical long strings. Excluding such configurations demands the following bound on the external momenta of the $n$-point functions \cite{MO3}
\begin{equation}
\sum_{i=1}^n h_i < k+n-3\label{La45}
\end{equation}
Only $n$-point functions obeying (\ref{La45}) would actually make sense. As it can be easily checked, (\ref{La45}) is in general stronger than the bound coming from unitarity. In fact, unitarity would merely imply
\begin{equation}
\sum_{i=1}^n h_i \leq  \frac{n}{2}(k-1)\label{yesssta}.
\end{equation}
The remarkable fact is that, for $n>2$, (\ref{La45}) implies (\ref{yesssta}) only when $k\leq 3$. This observation is probably important in relation to the recent \cite{1911.00378}, as in there the worldsheet instantons seem to play an important role in the localization of the correlation functions.  
\begin{figure}
\ \ \ \  \ \ \ \includegraphics[width=5.7in]{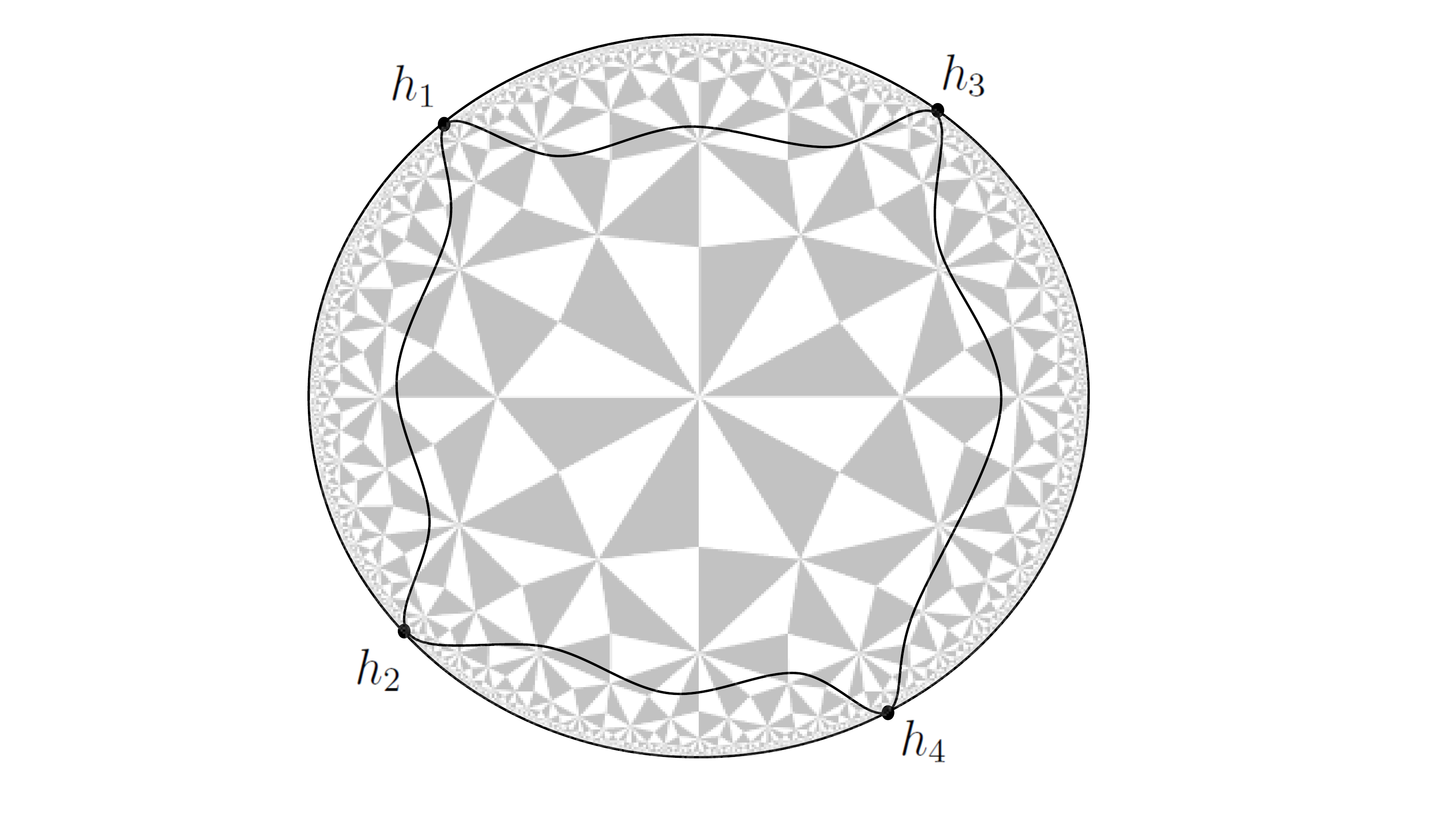}
\caption{Configuration associated to divergences at $\sum_{i=1}^4h_i=k+1$ in the $4$-point function.}
\label{Figure2}
\end{figure}

Let us finish this section with another feature of the $k=3$ theory that somehow is also related to the worldsheet instantons: $k=3$ is the fixed point of the Langlands duality, which in the case of affine $sl(2)_k$ symmetry corresponds to interchange \cite{Frenkel, Langlands, Langlands2}
\begin{equation}
k-2 \ \leftrightarrow \ \frac{1}{k-2}\, .
\end{equation}
From this point of view, it should probably not come to a surprise that at such a selfdual point the theory exhibits special features. One such feature is the coalescence of two marginal operators, which ultimately implies a special connection to the action of $c=25$ Liouville field theory, and consequently to $2D$ string theory. To see this, it is convenient to use again the representation (\ref{Waki}). In this representation, the worldsheet action can be thought of as a free CFT$_2$ perturbed with the marginal operator
\begin{equation}
\mathcal{O}_{(1,1)}=\lambda \ \beta\bar{\beta} e^{-\sqrt{\frac{2}{k-2}}\phi}.
\end{equation}
The theory, however, admits a second marginal operator; namely
\begin{equation}
 f_k\ \mathcal{O}_{(1,1)}^{k-2}=  f_k\ \lambda^{k-2} (\beta \bar{\beta } )^{k-2}e^{-\sqrt{{2}(k-2)}\phi } \, ,\label{Ooos1}
\end{equation}
where the exact coefficient $f_k$ was computed in \cite{GN3} and found to be
\begin{equation}
f_k = \pi^{k-3}\frac{\Gamma(3-k)}{\Gamma(k-2)}  \Big( \frac{\Gamma(\frac{1}{k-2})}{\Gamma(\frac{k-3}{k-2})}  \Big)^{k-2}.\label{Ooos2}
\end{equation}

Operators (\ref{Ooos1}) and (\ref{Ooos2}) are both exactly marginal and, in many respects, they are interchangeable. These operators, however, behave quite differently at large $k$: while operator $\mathcal{O}_{(1,1)}$ goes like $\sim e^{-\sqrt{2/k}\phi }$ and tends to 1 for finite $\phi $ in the semiclassical limit, operator $\mathcal{O}_{(1,1)}^{k-2}$ goes like $\sim e^{-\sqrt{2k}\phi }$ and thus vanishes in that regime; see \cite{Itzhaki1, Itzhaki2}. Thus, the latter operator describes non-perturbative effects in the worldsheet, i.e. the worldsheet instantons we discussed above \cite{Arash}. At finite $k$, in contrast, operators (\ref{Ooos1}) and (\ref{Ooos2}) compete, and at $k=3$ they actually coincide. This means that, as it typically happens with this class of confluent phenomena, a new marginal operator emerges at the special point; namely \cite{GL}
\begin{equation}
\tilde{\mathcal{O}}_{(1,1)} \equiv \lim _{k\to 3}\frac{\mathcal{O}_{(1,1)}+ f_k\, \mathcal{O}_{(1,1)}^{k-2}}{k-3}=(\text{const}\, + \log \beta\bar{\beta } -\sqrt {2}\phi ) e^{-\sqrt{2}\phi },\label{con}
\end{equation}
where we have used that $ f_{3} = -1$ (cf. $ f_{2} = 0$ and $ f_{k} = \infty $ for $k\in \mathbb{Z}_{>3}$). Then, integrating out the $\beta$-$\gamma$ system following the path integral techniques of \cite{HikidaSchomerus}, which also involves a shift in the $\phi $ direction, one can show that the correlation functions in the $SL(2,\mathbb{R})_{3}$ WZW model can always be written as correlation functions in the CFT$_2$ defined by the action\begin{equation}
S=\frac{1}{2\pi }\int_{\Sigma } d^2z \,\Big(\partial X \bar{\partial }X + \partial \varphi \bar{\partial }\varphi +2\pi \varphi \, e^{\sqrt{{2}}\varphi } +2\pi \, e^{\sqrt{{2}}\varphi } \Big),\label{Mc}
\end{equation}
which is, indeed, the complete action of $U(1)$ matter ($X$) coupled to $c=25$ Liouville field theory ($\varphi $). Notice the presence of the non-exponential marginal operator $\sim \varphi \, e^{\sqrt{2}\varphi}$ in (\ref{Mc}), and respectively in (\ref{con}). This is the type of operator that appears in the Lagrangian representation of the $2D$ string theories; see for instance \cite{McGreevy, KKK}. It is worth mentioning that exactly the same field redefinitions and functional integral techniques used in \cite{HikidaSchomerus} to prove the $H_3^+$ WZW-Liouville correspondence work here to show the relation between the confluent point $k=3$ and the $c=25$ theory, despite the {\it logarithmic} form of interaction (\ref{con}). 

\section{Superstrings and AdS$_3$/CFT$_2$}

Now, let us go back to the supersymmetric theory at the special point $k=1$ and discuss it in the context of the AdS$_3$/CFT$_2$ correspondence: We start by recalling that superstring theory admits NS-NS backgrounds of the form  
\begin{equation}
\text{AdS}_3 \times S^3 \times M_4
\end{equation}
with $M_4=T^4$ and $M_4=K_3$. These backgrounds are obtained in the near horizon limit of the NS$_1$/NS$_5$ system, where $k$ gives the number of NS$_5$ branes. Therefore, considering $k=1$ is actually considering the setup with a unit of NS flux. 

The theory also admits solutions of the form $M_4=S^3\times S^1$, but let us first focus on the case $M_4 = T^4$, which can be described as the WZW model on $SL(2,\mathbb{R})_k \times SU(2)_{k'} \times U(1) \times U(1) \times U(1) \times U(1)$. Supersymmetry demands the Kac-Moody levels of the $SL(2,\mathbb{R})_{k}$ and the $SU(2)_{k'}$ parts to be equal, and thus one gets
\begin{equation}
k=k'\ \ , \ \ \ c= \frac{3(k+2)}{k}+\frac 32 +\frac{3(k-2)}{k}+\frac 32 + 4 +\frac 42 =15,
\end{equation}
where the order of the terms in the expression for the central charge has been chosen in a way that makes it explicit the origin of each contribution. In particular, the third and the fourth terms in $c$ are the contribution coming from the $SU(2)_{k}$ piece, namely the superconformal field theory (SCFT) on $S^3$, which at $k=1$ is actually negative. This might seem puzzling; however, it has been argued that the theory can still be well-defined. 

The case $M_4 = S^3 \times S^1$ is different in this regard. It corresponds to the WZW model on $SL(2,\mathbb{R})_k \times SU(2)_{k'} \times SU(2)_{k''} \times U(1)$, and in this case supersymmetry relates the three levels of the curved pieces by a single equation, giving a 2-parameter family. One gets
\begin{equation}
k=\frac{k'k''}{k'+k''}\ \ , \ \ \ c= \frac{3(k+2)}{k}+\frac 32+\frac{3(k'-2)}{k'}+\frac 32+\frac{3(k''-2)}{k''}+\frac 32 + 1 +\frac 12 =15.
\end{equation}
Here, again, the order of the terms in $c$ makes it clear the origin of each contribution. Notice that now, when $k=1$, the levels of the two $3$-spheres can take the value $k'=k''=2$, so one has in the internal manifold a factor 
\begin{equation}\label{postalina}
 SU(2)_{2}\times SU(2)_{2},
\end{equation}
which yields a positive contribution to $c$.

In \cite{1803.04420}, the superstring theory on AdS$_3\times S^3\times S^3 \times S^1$ at $k=1$ was studied, and a simple CFT$_2$ model was proposed as its holographic dual. Such CFT$_2$ was constructed in terms of the free symmetric product orbifold. The Hilbert space of the symmetric product decomposes into twisted sectors, which are labeled by the conjugacy classes of the permutation group $S_N$. This permits to write down the spectrum and to verify that, at large $N$, it exactly reproduces the full spectrum of the long strings of the weakly coupled string theory, including states unprotected by supersymmetry. The correspondence is such that the spectral flow parameter $\omega $, representing the winding number of the string states, is mapped to the twisting number of corresponding states in the dual CFT$_2$. Specifically, the worldsheet theory considered in \cite{1803.04420} is the $\sigma$-model on AdS$_3\times {M}$, which yields the central charge
\begin{equation}
c=\frac{3(k+2)}{k}+\frac{3}{2}+c_{{M}} = 15.
\end{equation}
This implies that the contribution from the internal manifold at $k=1$ is $c_{{M}}=9/2$. Then, the SCFT on ${M}$ was taken to be the product of the SCFT on $S^1$ and 6 free fermions, and the proposed dual theory consists of a SCFT on the symmetric orbifold \cite{1803.04420}
\begin{equation}
 \big( c=1 \text{ CFT }\times \,{M}\big)^N\,/\,S_N\label{HJ}
\end{equation}
where the first factor in (\ref{HJ}) stands for a non-compact CFT$_2$ with central charge $c=1$. The presence of six free fermions can be understood as due to the contribution (\ref{postalina}), since the affine $su(2)_2$ current algebra admits a representation in terms of three free fermions.

The spectrum of superstrings on AdS$_3 \times M_7$ with pure NS-NS flux at $k=1$ was also analyzed in \cite{1803.04423}. For $M_7=S^3 \times S^3 \times S^1$ and $M_7=S^3 \times T^4$, it was argued there that there is a special set of physical states which agree in precise detail with the single particle spectrum of a free symmetric product orbifold. The theory on AdS$_3 \times S^3 \times S^3 \times S^1$ at $k=1$ was revisited in \cite{1904.01585}, where it was shown that the spacetime spectrum and the algebra of operators match those of the symmetric orbifold of $S^3 \times S^1$ in the large $N$ limit. 

In the last two years there have been many, very interesting developments in the realization of AdS$_3$/CFT$_2$ in presence of NS-NS fluxes at finite $k$: In \cite{1812.01007}, for example, superstring theory on AdS$_3 \times S^3 \times T^4$ at $k=1$ was described using the so-called hybrid formalism, in which the AdS$_3 \times S^3$ piece of the geometry is described in terms of the WZW model on the supergroup $PSU(1,1|2)_{k=1}$, which gives a model perfectly well defined at level $1$. Eberhardt, Gaberdiel, and Gopakumar showed in \cite{1812.01007} that, if the theory is described such a way, the string spectrum at $k=1$ does not exhibit the long string continuum, and it perfectly matches with the large $N$ limit of a symmetric product CFT$_2$. 

In reference \cite{1903.00421}, the authors argued that the CFT$_2$ dual of superstring theory on AdS$_3 \times S^3 \times T^4$ for generic NS-NS flux is actually the symmetric orbifold of the product between $\mathcal{N}=4$ super-Liouville and $T^4$; namely 
\begin{equation}
 \big( \mathcal{N}=4\, \text{Liouville} \times \,{T^4}\big)^N\,/\,S_N\, .\label{HJSS}
\end{equation}
At $k=1$, the Liouville factor disappears and theory reduces to the symmetric orbifold of $T^4$. Further evidence of the duality with the symmetric product based on Liouville theory was presented in \cite{1907.13144}, where the correlation functions were studied for the bosonic analogue of this duality, relating bosonic string theory on AdS$_3\times M$ to the symmetric orbifold of Liouville $\times M$. In \cite{1911.00378}, it was shown how to relate the $n$-point correlation functions of the supersymmetric theory on AdS$_3\times S^3 \times T^4$ at $k=1$ to those of the symmetric product CFT$_2$ on $T^4$. In particular, it was shown that the worldsheet correlation functions localize in the moduli space of worldsheet instantons, namely worldsheet configurations that cover the AdS$_3$ boundary holomorphically. As a consequence of this, the worldsheet correlators acquire the same structure as that of the symmetric orbifold correlators. In \cite{Oyi}, the theory at higher genus was studied and further evidence that the worldsheet correlators actually localize on such solutions was found, including $1/N$ corrections. All these results suggest the equivalence between the AdS$_3\times M_7$ string theory at $k=1$ and the spacetime orbifold CFT$_2$, raising hope in a precise realization of AdS$_3$/CFT$_2$ in the stringy regime.

%%%%%%%%%%%%%%%%%%%%%%%%%%%%%%%%%%

%%%%%%%%%%%%%%%%%%%%%%%%%%%%%%%%%%

%%%%%%%%%%%%%%%%%%%%%%%%%%%%%%%%%%

%%%%%%%%%%%%%%%%%%%%%%%%%%%%%%%%%%%%%%%%%%%%%%%%%%%%%%%

%%%%%%%%%%%%%%%%%%%%%%%%%%%%%%%%%%%%%%%%%%

\[\]    
The author is indebted to Chris Hull, Matt Kleban, Massimo Porrati, and Eliezer Rabinovici for many discussions and for an enjoyable collaboration in the subject. He also thanks the organizers of the Quantum Gravity in the Southern Cone VIII for the invitation to speak in such a nice conference. This work was partially supported by CONICET through the grant PIP 1109 (2017).


\begin{thebibliography}{666}



\bibitem{Verlinde} J.~de Boer, J.~Manschot, K.~Papadodimas and E.~Verlinde, 
``The Chiral ring of AdS(3)/CFT(2) and the attractor mechanism,''
  JHEP {\bf 0903}, 030 (2009)
 % doi:10.1088/1126-6708/2009/03/030
  [arXiv:0809.0507 [hep-th]].
  %%CITATION = doi:10.1088/1126-6708/2009/03/030;%%
 %55 citations counted in INSPIRE as of 20 Sep 2019

\bibitem{Pakman} A.~Dabholkar and A.~Pakman, 
``Exact chiral ring of AdS(3) / CFT(2),''
  Adv.\ Theor.\ Math.\ Phys.\ {\bf 13}, no. 2, 409 (2009)
 % doi:10.4310/ATMP.2009.v13.n2.a2
  [arXiv:hep-th/0703022].
  %%CITATION = doi:10.4310/ATMP.2009.v13.n2.a2;%%
 %62 citations counted in INSPIRE as of 20 Sep 2019
%\cite{Gaberdiel:2007vu}

\bibitem{Gaberdiel} M.~R.~Gaberdiel and I.~Kirsch, 
``Worldsheet correlators in AdS(3)/CFT(2),''
  JHEP {\bf 0704}, 050 (2007)
 % doi:10.1088/1126-6708/2007/04/050
  [arXiv:hep-th/0703001].
  %%CITATION = doi:10.1088/1126-6708/2007/04/050;%%
 %62 citations counted in INSPIRE as of 20 Sep 2019
%\cite{Giribet:2007wp}

%\cite{Pakman:2007hn}
\bibitem{Sever} 
  A.~Pakman and A.~Sever,
  ``Exact N=4 correlators of AdS(3)/CFT(2),''
  Phys.\ Lett.\ B {\bf 652}, 60 (2007)
  %doi:10.1016/j.physletb.2007.06.041
  [arXiv:0704.3040 [hep-th]].
  %%CITATION = doi:10.1016/j.physletb.2007.06.041;%%
  %42 citations counted in INSPIRE as of 20 Dec 2019

\bibitem{Rastelli} G.~Giribet, A.~Pakman and L.~Rastelli, 
``Spectral Flow in AdS(3)/CFT(2),''
  JHEP {\bf 0806}, 013 (2008)
  %doi:10.1088/1126-6708/2008/06/013
  [arXiv:0712.3046 [hep-th]].
  %%CITATION = doi:10.1088/1126-6708/2008/06/013;%%
 %32 citations counted in INSPIRE as of 20 Sep 2019%\cite{deBoer:2008ss}

\bibitem{Cardona}
  C.~A.~Cardona and C.~A.~Nunez,
  ``Three-point functions in superstring theory on AdS(3) x S3 x T4,''
  JHEP {\bf 0906}, 009 (2009)
  %doi:10.1088/1126-6708/2009/06/009
  [arXiv:0903.2001 [hep-th]].
  %%CITATION = doi:10.1088/1126-6708/2009/06/009;%%
  %17 citations counted in INSPIRE as of 20 Dec 2019
	
\bibitem{Gopakumar} R. Gopakumar, ``We have to talk more about string theory in AdS(3),'' talk delivered at the conference ``{20 Years Later: The Many Faces of AdS/CFT}'', at Princeton University, Oct. 31$^{\text{st}}$ - November 3$^{\text{rd}}$, 2017

\bibitem{Gaberdiel:2017oqg} M.~R.~Gaberdiel, R.~Gopakumar and C.~Hull,
``Stringy AdS$_{3}$ from the worldsheet,''
  JHEP {\bf 1707}, 090 (2017)
 % doi:10.1007/JHEP07(2017)090
  [arXiv:1704.08665 [hep-th]].
  %%CITATION = doi:10.1007/JHEP07(2017)090;%%
 %20 citations counted in INSPIRE as of 20 Sep 2019

\bibitem{1704.08667} K.~Ferreira, M.~R.~Gaberdiel and J.~I.~Jottar, 
``Higher spins on AdS$_{3}$ from the worldsheet,''
  JHEP {\bf 1707}, 131 (2017)
 % doi:10.1007/JHEP07(2017)131
  [arXiv:1704.08667 [hep-th]].
  %%CITATION = doi:10.1007/JHEP07(2017)131;%%
 %30 citations counted in INSPIRE as of 20 Sep 2019

\bibitem{1803.04420} G.~Giribet, C.~Hull, M.~Kleban, M.~Porrati and E.~Rabinovici, ``Superstrings on AdS$_{3}$ at ${k} =1$,''
  JHEP {\bf 1808}, 204 (2018)
  %doi:10.1007/JHEP08(2018)204
  [arXiv:1803.04420 [hep-th]].
  %%CITATION = doi:10.1007/JHEP08(2018)204;%%
 %30 citations counted in INSPIRE as of 20 Sep 2019

\bibitem{1803.04423} M.~R.~Gaberdiel and R.~Gopakumar, ``Tensionless string spectra on AdS$_{3}$,''
  JHEP {\bf 1805}, 085 (2018)
  %doi:10.1007/JHEP05(2018)085
  [arXiv:1803.04423 [hep-th]].  %%CITATION = doi:10.1007/JHEP05(2018)085;%%
  %37 citations counted in INSPIRE as of 20 Sep 2019

\bibitem{1812.01007} L.~Eberhardt, M.~R.~Gaberdiel and R.~Gopakumar, ``The Worldsheet Dual of the Symmetric Product CFT,''
  JHEP {\bf 1904}, 103 (2019)
 % doi:10.1007/JHEP04(2019)103
  [arXiv:1812.01007 [hep-th]].
  %%CITATION = doi:10.1007/JHEP04(2019)103;%%
 %18 citations counted in INSPIRE as of 20 Sep 2019

\bibitem{1904.01585} L.~Eberhardt and M.~R.~Gaberdiel, ``Strings on $\text{AdS}_3 \times \text{S}^3 \times \text{S}^3 \times \text{S}^1$,''
  JHEP {\bf 1906}, 035 (2019)
  %doi:10.1007/JHEP06(2019)035
  [arXiv:1904.01585 [hep-th]].
  %%CITATION = doi:10.1007/JHEP06(2019)035;%%
 %4 citations counted in INSPIRE as of 20 Sep 2019



\bibitem{1903.00421} L.~Eberhardt and M.~R.~Gaberdiel, ``String theory on $\boldsymbol{\text{AdS}_{\mathbf{3}}}$ and the symmetric orbifold of Liouville theory,''
  [arXiv:1903.00421 [hep-th]].
  %%CITATION = ARXIV:1903.00421;%%
 %8 citations counted in INSPIRE as of 20 Sep 2019

\bibitem{1907.13144} A.~Dei, L.~Eberhardt and M.~R.~Gaberdiel, 
``Three-point functions in AdS$_3$/CFT$_2$ holography,''
  [arXiv:1907.13144 [hep-th]].
  %%CITATION = ARXIV:1907.13144;%%
 %2 citations counted in INSPIRE as of 20 Sep 2019
%\cite{Maldacena:2000hw}

%\cite{Raeymaekers:2019dkc}
\bibitem{otros} 
  J.~Raeymaekers,
  ``On tensionless string field theory in AdS$_3$,''
  JHEP {\bf 1907}, 019 (2019)
  %doi:10.1007/JHEP07(2019)019
  [arXiv:1903.09647 [hep-th]].
  %%CITATION = doi:10.1007/JHEP07(2019)019;%%

\bibitem{otros2} 
  L.~V.~Eberhardt,
  ``Strings on AdS3,'' PhD Thesis, ETH Zurich (2019).
%  doi:10.3929/ethz-b-000385691
  %%CITATION = doi:10.3929/ethz-b-000385691;%%

\bibitem{MO1} J.~M.~Maldacena and H.~Ooguri, 
``Strings in AdS(3) and SL(2,R) WZW model 1.: The Spectrum,''
  J.\ Math.\ Phys.\ {\bf 42}, 2929 (2001)
 % doi:10.1063/1.1377273
  [arXiv:hep-th/0001053].
  %%CITATION = doi:10.1063/1.1377273;%%
 %423 citations counted in INSPIRE as of 20 Sep 2019
%\cite{Maldacena:2000kv}

\bibitem{MO2} J.~M.~Maldacena, H.~Ooguri and J.~Son, 
``Strings in AdS(3) and the SL(2,R) WZW model. Part 2. Euclidean black hole,''
  J.\ Math.\ Phys.\ {\bf 42}, 2961 (2001)
 % doi:10.1063/1.1377039
  [arXiv:hep-th/0005183].
  %%CITATION = doi:10.1063/1.1377039;%%
 %212 citations counted in INSPIRE as of 20 Sep 2019%\cite{Maldacena:2001km}

\bibitem{Frenkel} E.~Frenkel, 
``Lectures on the Langlands program and conformal field theory,''
 % doi:10.1007/978-3-540-30308-4_11
  [arXiv:hep-th/0512172].
  %%CITATION = doi:10.1007/978-3-540-30308-4_11;%%
 %76 citations counted in INSPIRE as of 20 Sep 2019
%\cite{Hikida:2007tq}

\bibitem{Nakayama} G.~Giribet and Y.~Nakayama, 
``The Stoyanovsky-Ribault-Teschner map and string scattering amplitudes,''
  Int.\ J.\ Mod.\ Phys.\ A {\bf 21}, 4003 (2006)
 % doi:10.1142/S0217751X06031697
  [arXiv:hep-th/0505203].
  %%CITATION = doi:10.1142/S0217751X06031697;%%
 %18 citations counted in INSPIRE as of 20 Sep 2019
%\cite{Giribet:2001ft}

%\cite{Giribet:2008ix}
\bibitem{Langlands} 
  G.~Giribet, Y.~Nakayama and L.~Nicolas,
  ``Langlands duality in Liouville-H(+)-3 WZNW correspondence,''
  Int.\ J.\ Mod.\ Phys.\ A {\bf 24}, 3137 (2009)
  %doi:10.1142/S0217751X09044607
  [arXiv:0805.1254 [hep-th]].
  %%CITATION = doi:10.1142/S0217751X09044607;%%
  %12 citations counted in INSPIRE as of 03 Dec 2019
	
	%\cite{Teschner:2010je}
\bibitem{Langlands2} 
  J.~Teschner,
  ``Quantization of the Hitchin moduli spaces, Liouville theory, and the geometric Langlands correspondence I,''
  Adv.\ Theor.\ Math.\ Phys.\  {\bf 15}, no. 2, 471 (2011)
  %doi:10.4310/ATMP.2011.v15.n2.a6
  [arXiv:1005.2846 [hep-th]].
  %%CITATION = doi:10.4310/ATMP.2011.v15.n2.a6;%%
  %105 citations counted in INSPIRE as of 03 Dec 2019
  
\bibitem{KS} D.~Kutasov and N.~Seiberg, 
``More comments on string theory on AdS(3),''
  JHEP {\bf 9904}, 008 (1999)
 % doi:10.1088/1126-6708/1999/04/008
  [arXiv:hep-th/9903219].
  %%CITATION = doi:10.1088/1126-6708/1999/04/008;%%
 %210 citations counted in INSPIRE as of 20 Sep 2019
%\cite{deBoer:1998gyt}

	%\cite{Banados:1992wn}
\bibitem{BTZ} 
  M.~Banados, C.~Teitelboim and J.~Zanelli,
  ``The Black hole in three-dimensional space-time,''
  Phys.\ Rev.\ Lett.\  {\bf 69}, 1849 (1992)
  %doi:10.1103/PhysRevLett.69.1849
  [hep-th/9204099].
  %%CITATION = doi:10.1103/PhysRevLett.69.1849;%%
  %2600 citations counted in INSPIRE as of 07 Jan 2020

\bibitem{Giveon:2005mi} A.~Giveon, D.~Kutasov, E.~Rabinovici and A.~Sever, 
``Phases of quantum gravity in AdS(3) and linear dilaton backgrounds,''
  Nucl.\ Phys.\ B {\bf 719}, 3 (2005)
 % doi:10.1016/j.nuclphysb.2005.04.015
  [arXiv:hep-th/0503121].
  %%CITATION = doi:10.1016/j.nuclphysb.2005.04.015;%%
 %62 citations counted in INSPIRE as of 20 Sep 2019

\bibitem{GKS} A.~Giveon, D.~Kutasov and N.~Seiberg, 
``Comments on string theory on AdS(3),''
  Adv.\ Theor.\ Math.\ Phys.\ {\bf 2}, 733 (1998)
 % doi:10.4310/ATMP.1998.v2.n4.a3
  [arXiv:hep-th/9806194].
  %%CITATION = doi:10.4310/ATMP.1998.v2.n4.a3;%%
 %379 citations counted in INSPIRE as of 20 Sep 2019
%\cite{Kutasov:1999xu}

\bibitem{Ooguri} J.~de Boer, H.~Ooguri, H.~Robins and J.~Tannenhauser, ``String theory on AdS(3),''
  JHEP {\bf 9812}, 026 (1998)
 % doi:10.1088/1126-6708/1998/12/026
  [arXiv:hep-th/9812046].
  %%CITATION = doi:10.1088/1126-6708/1998/12/026;%%
 %171 citations counted in INSPIRE as of 20 Sep 2019

\bibitem{SeibergWitten} N.~Seiberg and E.~Witten, 
``The D1 / D5 system and singular CFT,''
  JHEP {\bf 9904}, 017 (1999)
 % doi:10.1088/1126-6708/1999/04/017
  [arXiv:hep-th/9903224].
  %%CITATION = doi:10.1088/1126-6708/1999/04/017;%%
 %424 citations counted in INSPIRE as of 20 Sep 

\bibitem{RibaultTeschner} S.~Ribault and J.~Teschner, 
``H+(3)-WZNW correlators from Liouville theory,''
  JHEP {\bf 0506}, 014 (2005)
 % doi:10.1088/1126-6708/2005/06/014
  [arXiv:hep-th/0502048].
  %%CITATION = doi:10.1088/1126-6708/2005/06/014;%%
 %74 citations counted in INSPIRE as of 20 Sep 2019
%\cite{Giribet:2005ix}

\bibitem{HikidaSchomerus} Y.~Hikida and V.~Schomerus, 
``H+(3) WZNW model from Liouville field theory,''
  JHEP {\bf 0710}, 064 (2007)
  %doi:10.1088/1126-6708/2007/10/064
  [arXiv:0706.1030 [hep-th]].
  %%CITATION = doi:10.1088/1126-6708/2007/10/064;%%
 %45 citations counted in INSPIRE as of 20 Sep 2019
%\cite{Ribault:2005wp}

\bibitem{MO3} J.~M.~Maldacena and H.~Ooguri, 
``Strings in AdS(3) and the SL(2,R) WZW model. Part 3. Correlation functions,''
  Phys.\ Rev.\ D {\bf 65}, 106006 (2002)
 % doi:10.1103/PhysRevD.65.106006
  [arXiv:hep-th/0111180].
  %%CITATION = doi:10.1103/PhysRevD.65.106006;%%
 %211 citations counted in INSPIRE as of 20 Sep 2019%\cite{Giribet:2004zd}

\bibitem{GN3} G.~Giribet and C.~A.~Nunez, 
``Correlators in AdS(3) string theory,''
  JHEP {\bf 0106}, 010 (2001)
 % doi:10.1088/1126-6708/2001/06/010
  [arXiv:hep-th/0105200].
  %%CITATION = doi:10.1088/1126-6708/2001/06/010;%%
 %70 citations counted in INSPIRE as of 20 Sep 2019
%\cite{Dabholkar:2007ey}


\bibitem{GL} G.~E.~Giribet and D.~E.~Lopez-Fogliani, 
``Remarks on free field realization of SL(2,R)(k)/U(1) x U(1) WZNW model,''
  JHEP {\bf 0406}, 026 (2004)
  %doi:10.1088/1126-6708/2004/06/026
  [arXiv:hep-th/0404231].
  %%CITATION = doi:10.1088/1126-6708/2004/06/026;%%
 %19 citations counted in INSPIRE as of 20 Sep 2019%\cite{Giveon:1998ns}


%\cite{Giveon:2019gfk}
\bibitem{Itzhaki1} 
  A.~Giveon and N.~Itzhaki,
  ``Stringy Black Hole Interiors,''
  JHEP {\bf 1911}, 014 (2019)
%  doi:10.1007/JHEP11(2019)014
  [arXiv:1908.05000 [hep-th]].
  %%CITATION = doi:10.1007/JHEP11(2019)014;%%
  %1 citations counted in INSPIRE as of 10 Jan 2020
	
	%\cite{Itzhaki:2019cgg}
\bibitem{Itzhaki2} 
  A.~Giveon and N.~Itzhaki,
  ``Stringy Information and Black Holes,''
  [arXiv:1912.06538 [hep-th]].
  %%CITATION = ARXIV:1912.06538;%%
	
	%\cite{Giribet:2015kca}
\bibitem{Arash} 
  G.~Giribet and A.~Ranjbar,
  ``Screening Stringy Horizons,''
  Eur.\ Phys.\ J.\ C {\bf 75}, no. 10, 490 (2015)
%  doi:10.1140/epjc/s10052-015-3714-0
  [arXiv:1504.05044 [hep-th]].
  %%CITATION = doi:10.1140/epjc/s10052-015-3714-0;%%
  %9 citations counted in INSPIRE as of 10 Jan 2020


	%\cite{McGreevy:2003ep}
\bibitem{McGreevy} 
  J.~McGreevy, J.~Teschner and H.~L.~Verlinde,
  ``Classical and quantum D-branes in 2-D string theory,''
  JHEP {\bf 0401}, 039 (2004)
  %doi:10.1088/1126-6708/2004/01/039
  [hep-th/0305194].
  %%CITATION = doi:10.1088/1126-6708/2004/01/039;%%
  %130 citations counted in INSPIRE as of 07 Jan 2020

	%\cite{Kazakov:2000pm}
\bibitem{KKK} 
  V.~Kazakov, I.~K.~Kostov and D.~Kutasov,
  ``A Matrix model for the two-dimensional black hole,''
  Nucl.\ Phys.\ B {\bf 622}, 141 (2002)
  %doi:10.1016/S0550-3213(01)00606-X
  [hep-th/0101011].
  %%CITATION = doi:10.1016/S0550-3213(01)00606-X;%%
  %194 citations counted in INSPIRE as of 07 Jan 2020



%\cite{Eberhardt:2019ywk}
\bibitem{1911.00378} 
  L.~Eberhardt, M.~R.~Gaberdiel and R.~Gopakumar,
  ``Deriving the $\text{AdS}_{3}/\text{CFT}_{2}$ Correspondence,''
  [arXiv:1911.00378 [hep-th]].
  %%CITATION = ARXIV:1911.00378;%%

\bibitem{Oyi} 
  L. Eberhardt
  ``AdS$_3$/CFT$_2$  at higher genus,''
  [arXiv:2002.11729 [hep-th]].

























%%%%%%%%%%%%%%%%%%%%%%%%%%%%%%%%%%%%%%%
















































%\cite{McGreevy:2003ep}
%\bibitem{McGreevy:2003ep} 
%  J.~McGreevy, J.~Teschner and H.~L.~Verlinde,
%  ``Classical and quantum D-branes in 2-D string theory,''
%  JHEP {\bf 0401}, 039 (2004)
  %doi:10.1088/1126-6708/2004/01/039
%  [arXiv:hep-th/0305194].
  %%CITATION = doi:10.1088/1126-6708/2004/01/039;%%
  %130 citations counted in INSPIRE as of 28 Nov 2019


	
	

	

	
	
  \end{thebibliography}
  \end{document}